\documentclass[sigplan]{acmart}
\renewcommand\footnotetextcopyrightpermission[1]{}
\makeatletter
\newcommand{\atomicity}{\textsc{IntersectionAtomicity }}
\newcommand{\consistency}{\textsc{IntersectionCC }}

\newcommand{\visibleat}{\mathrm{\rightsquigarrow}}
\newcommand{\preceeds}{\mathrm{\xrightarrow{hb}}}
\newcommand\incircbin
{%
  \mathpalette\@incircbin
}
\newcommand\@incircbin[2]
{%
  \mathbin%
  {%
    \ooalign{\hidewidth$#1#2$\hidewidth\crcr$#1\bigcirc$}%
  }%
}
\newcommand{\notinsub}{\incircbin{-}}
\newcommand{\insub}{\incircbin{\land}}

\newcommand{\localsub}{\textsc{LSet }}
\newcommand{\remotesub}{\textsc{RSet }}

\makeatother

\usepackage{amsmath}
\usepackage{nccmath}
\usepackage{hyperref}
\usepackage{tikz}
\usepackage{fancyvrb}

\acmConference{arxiv}{2026}{Online}

\title{A framework to reason about consistency and atomicity guarantees in a sparsely-connected, partially-replicated peer-to-peer system}
\author{Sreeja S. Nair}
\affiliation{%
  \institution{DittoLive Inc.}
  \country{France}
}
\author{Nicholas E. Marino}
\authornote{The work was done while at DittoLive Inc.}
\affiliation{%
  \institution{DittoLive Inc.}
  \country{USA}
}
\author{Nick Pascucci}
\affiliation{%
  \institution{DittoLive Inc.}
  \country{UK}
}
\author{Russell Brown}
\authornotemark[1]
\affiliation{%
  \institution{DittoLive Inc.}
  \country{UK}
}
\author{Arthur P. R. Silva}
\affiliation{%
  \institution{DittoLive Inc.}
  \country{Sweden}
}
\author{Tim Cummings}
\affiliation{%
  \institution{DittoLive Inc.}
  \country{USA}
}
\author{Connor M. Power}
\affiliation{%
  \institution{DittoLive Inc.}
  \country{USA}
}

\begin{document}

\begin{abstract}
For an offline-first collaborative application to operate in true peer-to-peer fashion, its collaborative features must function even in environments where internet connectivity is limited or unavailable.
Each peer may only be interested in a subset of the application data relevant to its workload, and this subset can overlap in different ways with those of other peers. 
Limitations imposed by access control and mesh network technologies often result in peers being sparsely connected. 
Reasoning about consistency in these systems is hard, especially when considering transactional updates that may alter different sets of data in the same transaction.
We present \atomicity and \consistency as models to reason about offline-first collaborative applications that are sparsely-connected and rely on partially replicating different subsets of a broader set of data.
We then use these models to propose a set of guidelines to help developers design their application with atomicity and consistency guarantees.

\end{abstract}

\maketitle

\section{Introduction}
\label{sec:intro}

Historically, merging and syncing collaborative data changes has been a non-trivial problem.
The Operational Transformation algorithm\cite{operational-transform} used by popular applications like Google Docs\cite{google-docs-ot} and Apache Wave\cite{wave-ot} requires a centralized server to coordinate changes.
As such, these systems provide only limited offline capabilities.
Users can continue to edit documents while offline, but they must reconnect to the server in order to resynchronize and merge their local data with changes from other users.

More recently, specialized data types known as \textit{conflict-free replicated data types} (CRDTs) \cite{crdts} have gained popularity as a way to allow merging data changes between peers without necessarily requiring a centralized service to do so.
Despite the theoretical promise, however, many real-world implementations of CRDTs still require centralized services to merge and synchronize changes. Examples include popular databases such as Riak\cite{riak-crdts}, Redis\cite{redis-crdts}, and Cosmos DB\cite{cosmos-crdts}.
This tendency to still require some central coordination despite the potential may partly stem from limitations in older types of CRDTs.
Of the implementations described in \citet{crdts}, state-based CRDTs require sending large amounts of data over the network, and operation-based CRDTs require network guarantees that can be difficult to uphold in systems with sparse/intermittent connectivity.
Neither are suitable for applications using peer-to-peer mesh technologies, such as Bluetooth, which have comparatively low bandwidth.
While newer techniques like $\Delta$-CRDTs\cite{delta-crdts} can overcome these limitations, challenges remain in understanding the types of consistency guarantees we can provide when building fully peer-to-peer synchronization features. This is especially true for applications that require support for transactional updates that operate on multiple subsets of the application data.

Even with the gains in mobile connectivity over recent decades, many real-world applications could benefit from the ability to synchronize data directly between users, without needing access to a remote server.
Cellular internet is not always reliable or ubiquitous—even Wi-Fi networks, which are often considered more dependable and widespread, have dead zones in large facilities like warehouses or manufacturing plants.
Furthermore, modern mobile devices typically already have the capability to communicate directly with each other by way of Bluetooth or Wi-Fi Direct\cite{wifi-direct}, so deploying these kinds of features is not limited by the available hardware.
Instead it is the state of software techniques and lack of models to reason about peer-to-peer replication that pose challenges.
By addressing some of the software challenges, we hope to facilitate the creation of more offline-first peer-to-peer applications, and thereby help to eliminate the limitations that centralized client-server designs impose.

In this paper, we begin by illustrating some of the typical challenges faced by such offline-first applications (\autoref{sec:problem}).
We then define the system model (\autoref{sec:model}) for a data synchronization system with devices that have a defined interest set, with an aim toward running peer-to-peer collaborative applications on a sparsely connected mesh network of devices.
For this system model, we study the trade-offs involved in providing traditional atomicity and consistency guarantees (\autoref{sec:tradeoffs}). 
We introduce two models, \atomicity and \consistency, to reason about a peer-to-peer system where peers subscribe to a restricted set of data, and also discuss the conditions under which they are upheld (\autoref{sec:guarantees}).

\section{Problem Space} 
\label{sec:problem}

Consider an application designed for use by an aircraft maintenance crew working at different locations within a hangar, where connectivity may be unavailable. 
Given this, a crew supervisor and the crew members must coordinate between them to work collaboratively.

In this example scenario, let Alice and Bob be aircraft maintenance workers, with Alice tasked with maintaining the aircraft's exterior paint and Bob with the landing gear. 
Let Sarah be the crew supervisor. 
Although all three workers have a mobile device, the data required for each worker to perform their specific job may be distinct according to their role.

For instance, when an aircraft arrives for maintenance, Sarah's device must be updated with all the tasks that must be performed for the given aircraft.
When Alice and Bob start their shift, their devices sync with either Sarah's device or with a stationary workstation, which, courtesy of a persistent wired connection, always has all the data necessary for Alice and Bob to complete the tasks assigned for their shifts.
Alice will subscribe to painting tasks and Bob will subscribe to tasks related to the landing gear.
Note that if there is a task that requires some painting work on the landing gear, that task will be present on \emph{both} Alice's and Bob's devices.
Alice and Bob can independently work on their respective tasks and the updates will be opportunistically replicated between them, Sarah, and the stationary workstation.

Offline-first collaborative applications, similar to the kind described in the example scenario, require a  replication protocol developed to ensure the integrity of the data subscribed to by each device.
The developer needs a guarantee model to aid their reasoning about the data integrity across different devices.
Existing literature suggests that the highest consistency level achievable while maintaining high availability is Transactional Causal+ Consistency (TCC+)\cite{cure-protocol}.
TCC+ comprises of atomicity, causal consistency, and strong convergence guarantees.
Atomicity ensures \emph{all-or-nothing} semantics of an application, causal consistency helps the developer preserve the happens-before relationship, and convergence ensures that the different copies of the same data converge to the same state.

To demonstrate the three property guarantees, let us revisit the scenario from the previous example. 
Imagine that Bob, one of the maintenance workers, is assigned the maintenance checklist task of replacing a bolt in the landing gear. 
To complete this task, Bob must first obtain a new bolt from the inventory and then proceed with the replacement. 
In the context of the application, this transaction would be represented as follows: 
\begin{Verbatim}[fontsize=\small]
transaction replace_bolt {
    inventory.bolts.new -= 1;
    landing_gear.bolt.replaced_on = '2024-02-16';
    inventory.bolts.old += 1;
    checklist.landing_gear.bolt.health = true;
}
\end{Verbatim}
Just like any other connected peer subscribed to the transaction, the device belonging to Sarah, the crew supervisor, must view Bob's task as a single, indivisible unit. 
Without this atomicity guarantee, discrepancies in peer data are probable.

Now consider that Alice, the other maintenance worker, is assigned the task of painting the bolt once it has been replaced by Bob. 
To complete this task, Alice needs to first ensure that the new bolt is in place. 
Once confirmed, Alice can proceed to obtain the white paint from inventory and then paint the bolt. 
In the application, this transaction would be represented as follows:
\begin{Verbatim}[fontsize=\small]
transaction paint_bolt {
    inventory.paint.white -= 1;
    landing_gear.bolt.painted_on = '2024-02-16';
    checklist.landing_gear.bolt.paint = true;
}
\end{Verbatim}
Once completed, Sarah receives the update indicating that Alice has finished painting the bolt. 
However, without the causality guarantee, Sarah may mistakenly believe that Alice painted the old bolt, as Bob's task occurred earlier in the causal past and Sarah would not have received the update.

Let's consider another scenario: as part of the landing gear checklist procedure, Bob notices a small crack in the paint of the landing gear tube and applies a patch to it. 
Represented as a transaction, this operation would be as follows:
\begin{Verbatim}[fontsize=\small]
transaction paint_tube {
    inventory.paint.white -= 2;
    landing_gear.tube.painted_on = '2024-02-16';
    checklist.landing_gear.tube.paint = true;
}
\end{Verbatim}
With the convergence guarantee, any device, irrespective of its observed ordering of updates, converge to the same state. 
For instance, if at the start of the day inventory indicates 20 liters of white paint, subscribing peers will view both the deductions of white paint and inventory state as follows:
\begin{Verbatim}[fontsize=\small]
inventory.paint.white = 17;
\end{Verbatim}

Let us now introduce David, the crew's maintenance checklist supervisor. 
Since David is only interested in (and subscribed to) updates related to the maintenance checklist, Alice and Bob replicate only the relevant information to David. 
Once received, David's device observes the transactions as follows:
\begin{Verbatim}[fontsize=\small]
transaction replace_bolt {
    checklist.landing_gear.bolt.health = true;
}
transaction paint_bolt {
    checklist.landing_gear.bolt.paint = true;
}
transaction paint_tube {
    checklist.landing_gear.tube.paint = true;
}
\end{Verbatim}
Despite having only partial information about the transactions, David has all the information to interact with other devices consistently and safely.
Traditional atomicity and consistency guarantees do not suffice in reasoning about partial replication in applications such as this example; we must ensure that we maintain the correct causal order of updates even when only subsets of a transaction's updates are replicated to a given device, \emph{and} those subsets must still be applied atomically.

\section{System Model}
\label{sec:model}
A \emph{node} is a physical device with some type of connectivity, local storage, and processing power.
A node may connect to one or more neighbouring nodes using any available technology/protocol.
These connections may be direct or routed, with the only constraint being that the nodes should be able to communicate in a peer-to-peer manner.
When two nodes connect, we call them \emph{peers} since there is no hierarchical relationship between them (as opposed to nodes with relationships like client/server, primary/secondary, and so on).

Nodes may be dispersed across a geographical area and may be mobile; i.e., they may change location at any time.
This mobility might affect the node's connectivity, rendering a previously connected node unreachable or forging new connections to nodes that were previously unconnected.
The node might also remain unconnected during or after this movement.
As a consequence, a connection to a desired node is not always guaranteed. 

Nodes may choose to subscribe to data relevant to their workflow. 
The union of a node's subscriptions is called its \emph{subscription set}.
This might comprise the entire data set or, more often, only a subset of it.
The node stores the data it is subscribed to locally which enables uninterrupted work (e.g. mutation of the data) even without connectivity. 
This makes the application offline-first: it may modify any data it has available locally, may insert new data which is stored locally on the node, and it  may replicate that data to other nodes which are interested in it.

Each node has a \emph{state} that consists of a set of objects.
The state can be updated by an \emph{operation}.
Multiple operations can be combined to be atomically applied, via a \emph{transaction}.
A transaction that originates at a given node can be called a \emph{local transaction} in the context of that node.
Any transactions that a node receives via replication can be called a \emph{remote transaction}.
Both local and remote transactions update the state of a given node.

When a node connects to a peer node, they first exchange a statement of their subscriptions (e.g. "all landing gear data") before any data exchange.
The peers subsequently exchange data that the other is interested in, ensuring that all relevant data is transmitted between the peers which matches their respective subscription sets.
While sending data to a connected peer we assume that the data is sent and received atomically.
This feature---called \emph{syncing at the edge}\cite{ditto}---enables the \emph{collaborative} aspects of the applications even in the absence of an internet connection.

In order to make this syncing system efficient the peers may also exchange a summary of the data they have together with their subscription information.
This summary might make use of causality tracking mechanisms such as \emph{version vectors}\cite{version-vectors} to detect situations where the other peer already has some/all of a given peer's data.
The peer receiving this information will compute the \emph{diffs} that it needs to send to the other peer to bring it up to date. 
Such a summary exchange is purely an optimization.
Note that any two peers can only be up to date with data that falls inside the intersection of their subscription sets, since each would have only the data that matches their own subscription set.

The data present on a node may be a superset of the data that node is subscribed to, if it contains data it does not request from other peers. 
In this case we assume the application developer is not expecting any guarantees about the data which falls outside its subscription set, hence we do not consider that data to uphold any guarantees.
It is possible that a remote peer is interested in data that the local node holds but is not subscribed to itself.
The remote peer will express this by communicating their \emph{subscription set} upon connection.
In this case, we still reason about guarantees upheld by this data in the context of the subscribed peer.
Taken together, we consider guarantees pertaining to the data at a node only if it falls inside its own subscription set.

In many applications, different nodes will have different roles, and hence their access may be restricted to only subsets of the application data.
For example, consider a scenario where the maintenance workers health information is also managed by the same application.
Sarah, their supervisor can potentially have access to this information in case of emergencies, but Alice and Bob, being coworkers, shouldn't be able to access each other's health information.
These restrictions are required to be upheld for security and confidentiality reasons.
We refer to the set of data the node is allowed to access as their \emph{permissions}.

For our purposes, permissions can be combined with subscriptions for any given node.
We define the \emph{interest set} to be the intersection of a node's subscriptions and permissions.
This ensures that a data is sent to a peer only if it matches its subscription \emph{and} permissions.

A node may change its interest set (permissions and/or subscriptions). 
It may also choose to clear all of its locally stored data and start fresh.
In both the cases, we consider that the present node leaves the system and a new node joins with the new configuration.

\section{Trade-offs for applications using an edge sync system}
\label{sec:tradeoffs}

The design of a replication protocol that supports offline-first collaborative applications involves three key dimensions:
\begin{itemize}
    \item The interest set of each individual node, which is the intersection of their subscriptions and permissions.
    \item The network topology and connectivity through which the nodes are interconnected.
    \item A set of guarantees that the application must maintain.
\end{itemize}

Let us examine these dimensions in detail.

\textbf{Interest set: } 
As explained in \autoref{sec:model}, rather than synchronizing all transactional data across distributed peers, an offline-first, peer-to-peer application  operates by synchronizing only the data that align with their respective interest sets.
The lower bound on the size of an interest set is determined by the data the application requires to enable seamless offline-first operations and collaboration.
Conversely, the upper bound is restricted by a node's permission set.

\textbf{Network topology and connectivity: } 
Depending on the type of application, the developer may or may not be able to influence the topology of the network.
At one end of the spectrum, the developer has the flexibility to choose static connections among nodes. 
In addition, the network layer may have 
a routing algorithm to ensure that specific peers will be connected.
In stark contrast, at the other end of the spectrum, the developer has no control over the connectivity of the peers and, consequently, no control over the network topology. 
In such cases, when peers are in close proximity, they harness compatible communication protocols to connect opportunistically.

\begin{figure}
\includegraphics[width=0.45\textwidth]{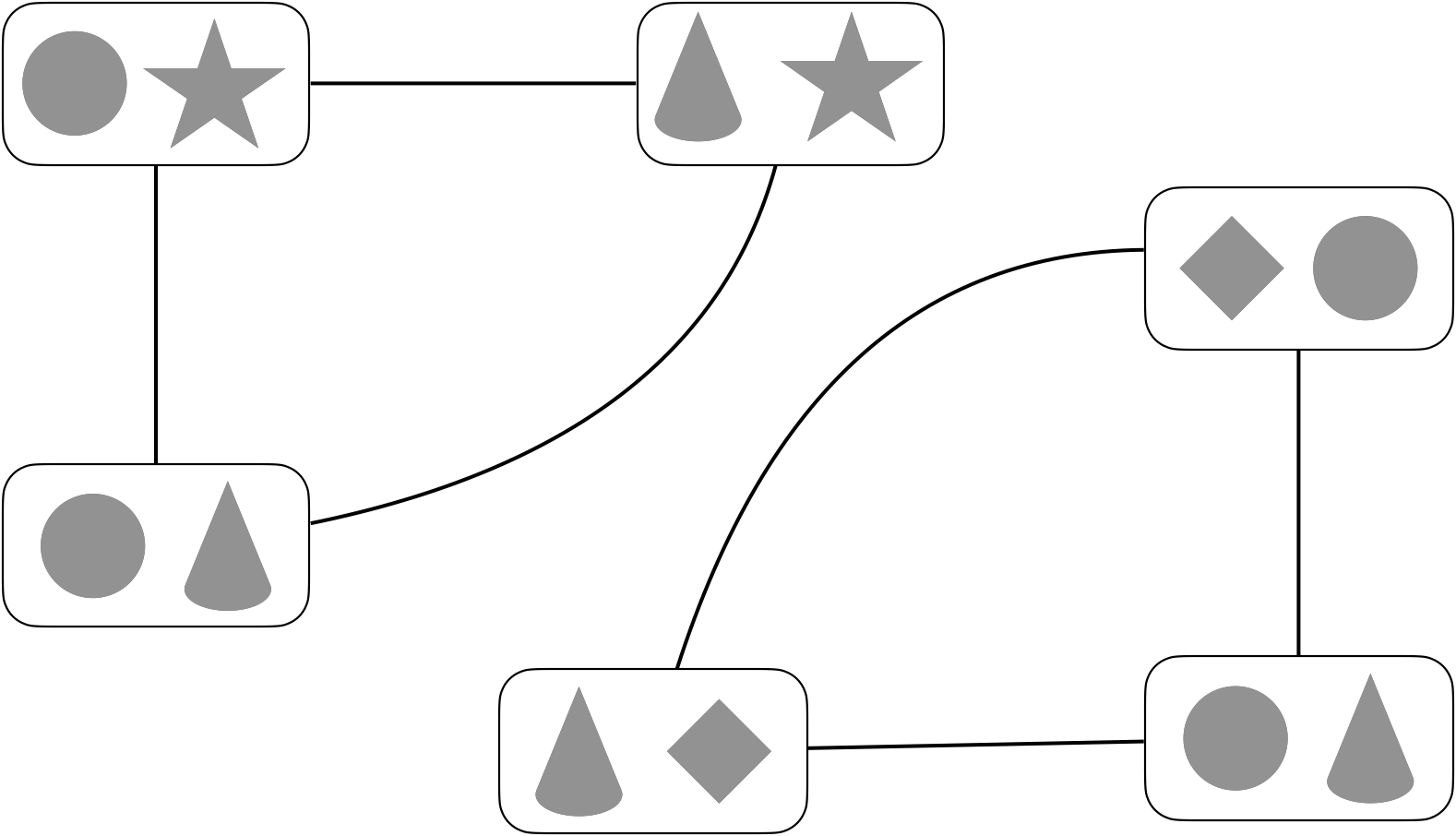}
\centering
\caption{A peer network graph consisting of multiple components, where \emph{interest sets} are satisfied, yet network partitions still exist. 
A peer is represented by a rectangle.
Connections between peers are indicated with arrows.
Different shapes represent different \emph{interest sets}.}
\label{fig:islands}
\end{figure}

All updates matching a node's interest set should eventually reach a node through its peers.
The network layer must ensure that each node can eventually connect to one or more peers that contain a superset of the node's entire interest set.
Additionally, the network layer should avoid the creation of \emph{data islands}. 
A \textit{data island} is defined as a connected subgraph of peers that lack connectivity to other peers with overlapping interest sets.

\autoref{fig:islands} depicts the formation of data islands among multiple peer groups.
The group located at the upper left include three nodes whose interest sets are satisfied by the peers they connect to. 
Since the interest set of each node in the group is covered, each node assumes they are receiving all the updates in their interest set.
However, the other group also has data that matches the interest sets of nodes in this group.
Since these groups are insufficiently connected, they form data islands.
Note that a single node cannot result in a data island since it can infer that it has to receive data from somewhere.
The number of nodes in the system and their interest sets are dynamic, which presents challenges in detecting these islands.

\textbf{Application guarantees: }
As we saw in \autoref{sec:problem}, collaborative applications benefit from atomicity, causal consistency, and convergence guarantees.
 
Following \citet{consistency-survey, burckhardt2014principles}, we model executions at a given node as an \emph{abstract execution} \[A = (H, vis, ar)\] where $H$ is a set of operations forming the node's \emph{history} and the visibility relation $vis$ combined with the arbitration $ar$ defines the relations between them.
Consider two operations $o_1$ and $o_2$. 
We write $o_1 \xrightarrow{vis} o_2$ to specify that $o_1$ was visible to $o_2$.
$ar$ is a total order of operations on the history that specifies how the system resolves conflicts due to concurrent operations.
The order in which a node executes operations is called the \emph{program order}. 
The \emph{happened-before} relation ($hb$) is the transitive closure of the union of visibility and program order.

If an operation $o$ is noted \emph{visible at}($\visibleat$) node $N$ at a given point in time, for any operation $o'$ that happens at $N$ later, $o \xrightarrow{vis} o'$.

Transactions group operations which must be applied together. 
Informally, atomicity guarantees are upheld when any given transaction is either atomically applied or not applied at all at any given node, i.e. it's an \emph{all-or-nothing} guarantee.
Formally, atomicity of a transaction $T$ can be defined as
\begin{equation}
Atomicity(T) \triangleq \forall N. \forall o_1, o_2 \in T. o_1 \visibleat N \implies o_2 \visibleat N     
\end{equation}
In other words, if operations $o_1$ and $o_2$ belong to a transaction $T$ and $o_1$ is visible at a given node $N$ in the system, then $o_2$ is also visible at the same node.

Causal consistency preserves the happens-before relationship for the updates, and strong convergence ensures that the nodes having the same set of updates converge to the same state.
Formally, causal consistency can be defined as
\begin{equation}
\begin{split}
CC \triangleq \forall N. \forall o_1, o_2 \in O. & o_1 \preceeds o_2 \wedge o_2 \visibleat N \implies o_1 \visibleat N
\end{split}
\end{equation}
In other words, if two operations are ordered by happened-before relationship, $o_1 \preceeds o_2$, and if $o_2$ is visible at a given node $N$, then $o_1$ should also be visible at the same node.

The strong convergence property states that two nodes observing the same set of updates converge to the same state\cite{consistency-survey}.
% \todo{define?}
In practice, strong convergence can be easily achieved by using Conflict-free Replicated Data Types\cite{crdts}.
When a node receives the state of a remote peer, it \emph{merges} the incoming state into its local state.
If the merge function is idempotent, associative, commutative and is the least upper bound of the two states, strong convergence is guaranteed\cite{crdts}.

Causal+ consistency combines causal consistency and strong convergence\cite{consistency-survey}.
Transaction Causal+ Consistency (TCC+) combines atomicity and causal+ consistency.

\subsection{Application guarantees and interest sets}
\label{subsec:guarantee-sub}
As discussed, a node's interest set depends on its role in the application (which includes the data it is subscribed to and the permissions of the node).
Let us discuss how the their definition impacts the guarantees.

When a node connects to a peer, the following cases represent the combinations of their interest sets:
\begin{enumerate}
    \item \label{enum:disjoint} the remote and local interest sets are disjoint
    \item \label{enum:same} the remote's interest set is the same as the local's
    \item \label{enum:super} the remote's interest set is a superset of the local's
    \item \label{enum:sub} the remote's interest set is a subset of the local's
    \item \label{enum:overlap} the remote's interest set overlaps that of the local's
\end{enumerate}

In the case of \autoref{enum:disjoint}, there is nothing to replicate and the session can close. 
In the case of \autoref{enum:same}, the local and remote peers are interested in the same data, and all guarantees that are held at the local node are also upheld at the remote.
The remaining cases are the most interesting ones.
In \autoref{enum:super}, for all the transactions that include only objects that match the local interest set, atomicity is maintained. 
Assuming that all the updates received from both local and remote origins are bundled in a single atomic replication update, causal consistency is preserved too.
In \autoref{enum:sub} atomicity and consistency are maintained since everything is a subset of what the local has.
In \autoref{enum:overlap}, for all transactions that include only the objects that match the intersection of the local interest set, atomicity and causal consistency is maintained. 
For the rest of the data that falls outside these conditions, eventual consistency and convergence are ensured\footnote{Provided the network ensures that the data eventually reaches the interested nodes.}.

Note that strong convergence requires two nodes observing the same set of update operations to reach the same state.
Since we require all the updates relevant to a node's interest set to eventually be delivered to the node, the use of CRDTs will trivially satisfy strong convergence guarantees for the partial view of the data.

Due to the inherent nature of the partial view of the data at each node, we see the need to impose some restrictions on the guarantees that a node in the system can satisfy.
In \autoref{sec:guarantees} we define these guarantees that each node must uphold locally with this partial view so that the entire system respects Transactional Causal+ Consistency(TCC+).

\subsection{Application guarantees and network topology}
\label{subsec:guarantee-network}

Let us discuss the relation between network topology and guarantees to address the case where the developer may define the network topology.

Consider a forest-like network topology where the edge nodes with the smallest interest set are situated at the leaves and the interest set widens as we approach the roots, culminating in the cloud that subscribes to everything.
In this hierarchical model, the transactions that originate in the leaf nodes maintain their atomicity and causality trivially since they are replicated to larger nodes. 
The transactions that originate in the root nodes may be chopped into several parts while transmitting to the nodes lower in the hierarchy.
Nevertheless, since the lower nodes are not interested in the remaining operations inside the transaction, the system as a whole can be considered as maintaining TCC+ guarantees.

The problematic scenario comes when the path between two nodes contain another node (or a set of nodes) with a smaller interest set.
Consider three nodes $N_1$, $N_2$ and $N_3$.
Assume that $N_1$ and $N_3$ have interest set $s_1 \cup s_2$ and is connected through $N_2$ with interest set $s_2$.
If a transaction originating at $N_1$ mutates data in both $s_1$ and $s_2$, while replicating this transaction to $N_2$, the transaction will be narrowed down to the mutations in $s_2$.
Now, when $N_2$ replicates with $N_3$, it results in a partial knowledge for $N_3$, breaking the atomicity guarantees.
The same reasoning would also apply to consistency guarantees.

\section{Guarantees for designing offline-first collaborative applications}
\label{sec:guarantees}

\begin{table*}[ht]
\begin{tabular}{c|c c c c}
\toprule
 & \localsub $\subset$ \remotesub & \localsub $\supset$ \remotesub & \localsub $=$ \remotesub & \localsub $\cap$ \remotesub $\neq$ $\emptyset$ \\
 % \localsub $\cap$ \remotesub $=$ $\emptyset$\\
\hline
\atomicity of a local transaction & \checkmark & \checkmark & \checkmark & \localsub $\cap$ \remotesub \\
\atomicity of a remote transaction & \localsub & \checkmark & \checkmark & \localsub $\cap$ \remotesub \\
\consistency of a single object & \checkmark & \checkmark & \checkmark & \checkmark \\
\consistency of multiple objects & \localsub & \checkmark & \checkmark & \localsub $\cap$ \remotesub \\
Convergence & \checkmark & \checkmark & \checkmark & \checkmark \\
\bottomrule
\end{tabular}
\caption{\atomicity and \consistency guarantees compared to peer interest sets. 
Each cell contains the predicate signifying the set of data that upholds the guarantees. \checkmark indicates an always true predicate.}
\label{tab:subguarantees}
\end{table*}

As discussed in \autoref{subsec:guarantee-sub}, there are specific conditions that must be intact for a peer-to-peer synchronization session to maintain atomicity and causal consistency guarantees.
We formalize these conditions into \atomicity and \consistency.
Informally, a receiving node preserves \atomicity if it preserves atomicity for a given transaction for updates to the data that is in its interest set.
Put another way, the system preserves the \atomicity of a transaction if a node observing one update in the transaction observes all other updates in the same transaction for the updates modifying the data the node is subscribed to.
Similarly, a receiving node preserves \consistency if it preserves causal consistency for updates to the data in its interest set.
Developers benefit from understanding and applying these concepts in their applications: \atomicity enables transactional guarantees in subscription-based synchronization, while \consistency ensures their system maintains consistency across nodes.

Note that convergence is trivially preserved when the data types are inherently convergent, as in the case of CRDTs. 
This is because an operation serves as an atomic unit of update, capable of modifying only one data object at a given time.

We denote $S_N$ as the interest set of a given node $N$.
If an operation mutates a data that falls within the interest set $S_N$ , we write $o \insub S_N$ ; otherwise, we write $o \notinsub S_N$ .

Following the formalization we adopted from \citet{consistency-survey}, and discussed in \autoref{sec:tradeoffs}, \atomicity and \consistency can be defined as follows:

A transaction, T, maintains \atomicity if for any node N in the system
\begin{equation}
\begin{split}
\atomicity(T) \triangleq \\
\forall o_1, o_2 \in T. o_1 \visibleat N \implies o_2 \visibleat N \lor o_2 \notinsub N
\end{split}
\label{eq:atomicity}
\end{equation}

An operation is said to maintain \consistency if, an update is visible at a node, the entire \textit{causal history} of that update that matches the node's interest set is visible.
\begin{equation}
\begin{split}
\consistency \triangleq \\
\forall o_1, o_2 \in O. o_1 \preceeds o_2 \land o_2 \visibleat N \implies o_1 \visibleat N \lor o_1 \notinsub N
\end{split}
\label{eq:consistency}
\end{equation}

Now let us examine how these guarantees will be maintained in a peer-to-peer synchronization session.
We call \localsub the interest set of the local peer that sends the update, and \remotesub the interest set of the remote peer that receives the update.
\consistency of a single object (also called coherence) and convergence are trivially maintained with the use of CRDTs.
When \localsub is a superset or equals the \remotesub, \atomicity of local and remote transactions and \consistency across multiple objects are maintained.
Note that this is because the sending node has more information than the receiving node, so the receiving node receives all the information it needs.
When the sending peer might not have the full information; that is, when \localsub $\subset$ \remotesub and \localsub $\cap$ \remotesub $\neq$ $\emptyset$, the guarantees are restricted to the data that falls under the intersection of the interest sets of both the nodes.
\autoref{tab:subguarantees} summarizes the conditions under which the guarantees are preserved with respect to the relation between the peers' interest sets.

If all peer-to-peer connections in the system maintains \atomicity and \consistency guarantees along with the use of CRDTs, it is guaranteed that atomicity and causal+ consistency is maintained throughout the system.
This is because these conditions ensure that atomicity and causal consistency is maintained for the data that each node is interested in, thereby ensuring that any operation in the system maintains TCC+ guarantees globally.

\subsection{Upholding the guarantees}

If all nodes subscribe to all data, \atomicity and \consistency guarantees trivially ensure TCC+ guarantees in the system.
Since our system model is restricted, we have to make sure that these guarantees are maintained by individual nodes at all time.

If we have the flexibility to choose the topology of the network, we can ensure TCC+ in the system if the nodes form a hierarchical network with smaller subscriptions at the bottom and increasingly larger subscriptions towards the top.
Hierarchical replication models\cite{colony} that restricts the topology into a forest-like architecture with cloud nodes at the root and edge nodes as the leaves follow this model.

The developer may also choose to configure the interest sets of the nodes in the system in such a way to ensure TCC+ guarantees.
In this case, the interest sets need to be chosen in such a way that a transaction and all its causal history doesn't fall outside the intersection of any pair of interests sets in the system.
For example, consider a system with the following transactions
\begin{itemize}
    \item $T_1$ mutates set $s_1$ and causally depends on set $s_2$,
    \item $T_2$ mutates set $s_3$ and causally depends on the same set, and
    \item $T_3$ mutates sets $s_3$ and $s_4$ and causally depends on $s_4$.
\end{itemize}
The interest sets of the nodes can be $s_1 \cup s_2$, $s_3 \cup s_4$ or any superset of them.

Another approach is to replicate all the metadata across all nodes and replicate the actual data if it falls in the interest set of the node.\footnote{Optimized solutions that handle this case are proposed in literature\cite{practi}.}

\section{Related Work}
\label{sec:literature}
Reasoning about distributed systems is a challenge due to the complexity of the trade-off considerations involved.
\citet{consistency-survey} presented a taxonomy of non-transactional consistency models, extending the basic mechanisms defined by \citet{burckhardt2014principles} for defining eventual consistency models.
Both of these works, however, assume that the nodes in the system store a complete copy of the data. 
In addition, although \citet{partial-rep-consistency-formal} provides a framework to reason about the consistency guarantees, they assume disjoint interest sets.

\citet{cure-protocol} presents a replication protocol that provides TCC+.
The protocol considers the cloud environment where there are data centers equipped with multiple physical machines tasked with sharding and storing the data.
\citet{colony} extends these guarantees to the edge, but enforces a forest-like topology for the nodes.
\citet{practi} also tackles the problem of providing causal consistency at the edge by sending optimized metadata.
Both systems make the trade-offs we discussed, the former by manipulating the network topology and the latter by manipulating the "interest set" in some ways.

\section{Conclusion}
This paper introduces two models, \atomicity and \consistency, to reason about a peer-to-peer system where peers subscribe to a restricted set of data.
We show that if each peer preserves the guarantees provided by these models then the entire system upholds Transactional Causal+ Consistency, the highest possible guarantee that can be expected from a highly available distributed system without coordination.
We also list the conditions to be satisfied to uphold the guarantees for any given peer.
These models, together with their required conditions, enable developers to reason about consistency and atomicity guarantees in sparsely-connected, partially-replicated peer-to-peer systems.

\section*{Acknowledgements}
The authors would like to thank Toni Kemp for editorial support and Matheus Cardoso for all the interesting discussions.

\bibliographystyle{ACM-Reference-Format}
\bibliography{references.bib}

\end{document}